\documentclass{amsart}
\usepackage{xcolor}

\usepackage[hidelinks]{hyperref}

\usepackage{ccaption}

\usepackage{graphicx}
\usepackage{subfig}
\usepackage{makebox}

\calclayout

\usepackage{amsmath}

\usepackage{multicol}

\usepackage[backend=biber,style=nature]{biblatex}
\usepackage{tikz,xcolor,hyperref}

\definecolor{lime}{HTML}{A6CE39}
\DeclareRobustCommand{\orcidicon}{%
	\begin{tikzpicture}
	\draw[lime, fill=lime] (0,0) 
	circle [radius=0.16] 
	node[white] {{\fontfamily{qag}\selectfont \tiny ID}};
	\draw[white, fill=white] (-0.0625,0.095) 
	circle [radius=0.007];
	\end{tikzpicture}
	\hspace{-2mm}
}

\foreach \x in {A, ..., Z}{%
	\expandafter\xdef\csname orcid\x\endcsname{\noexpand\href{https://orcid.org/\csname orcidauthor\x\endcsname}{\noexpand\orcidicon}}
}

\begin{document}

\begin{titlepage}
    \begin{center}
       \vspace*{1cm}

        Enhanced monitoring of atmospheric methane from space over the Permian basin with hierarchical Bayesian inference  \\
        
        \vspace{1cm}

        \orcidA{}Clayton Roberts$^{1*}$, \orcidB{}Oliver Shorttle$^{1,2}$, \orcidC{}Kaisey Mandel$^{1,3,4}$, Matthew Jones$^{5}$,\\
        Rutger Ijzermans$^{5}$, \orcidE{}Bill Hirst$^{6}$ and \orcidD{}Philip Jonathan$^{7,8}$ \\
    
    \end{center}
   
    \vspace{1cm}
   
    \noindent $^{*}$ Corresponding author \\
    \noindent $^{1}$ Institute of Astronomy, University of Cambridge, Madingley Road, Cambridge CB3 0HA, UK\\
    \noindent $^{2}$ Department of Earth Sciences, University of Cambridge, Downing Street, Cambridge CB2 3EQ, UK \\
    \noindent $^{3}$ Statistical Laboratory, Department of Pure Mathematics and Mathematical Statistics, University of Cambridge, Wilberforce Road, Cambridge CB3 0WB, UK \\
    \noindent $^{4}$ The Alan Turing Institute, Euston Road, London NW1 2DB, UK\\
    \noindent $^{5}$ Shell Global Solutions International B.V., Grasweg 31, 1031 HW Amsterdam, the Netherlands\\
    \noindent $^{6}$ Atmospheric Monitoring Sciences, Amsterdam, the Netherlands\\
    \noindent $^{7}$ Shell Research Ltd., London SE1 7NA, UK\\
    \noindent $^{8}$ Department of Mathematics and Statistics, Lancaster University, LA1 4YW, UK
\end{titlepage}

\section*{Abstract}\label{sec:abstract}
Methane is a strong greenhouse gas, with a higher radiative forcing per unit mass and shorter atmospheric lifetime than carbon dioxide. The remote sensing of methane in regions of industrial activity is a key step toward the accurate monitoring of emissions that drive climate change. Whilst the TROPOspheric Monitoring Instrument (TROPOMI) on board the Sentinal-5P satellite is capable of providing daily global measurement of methane columns, data are often compromised by cloud cover. Here, we develop a statistical model which uses nitrogen dioxide concentration data from TROPOMI to efficiently predict values of methane columns, expanding the average daily spatial coverage of observations of the Permian basin from 16\% to 88\% in the year 2019. The addition of predicted methane abundances at locations where direct observations are not available will support inversion methods for estimating methane emission rates at shorter timescales than is currently possible.

\section{Introduction}\label{sec:introduction}
Methane and carbon dioxide are the two dominant anthropogenic greenhouse gases responsible for warming Earth’s atmosphere above its pre-industrial era temperature \cite{IPCC2013}. The current average atmospheric concentrations of these gases are $\approx$1,900 parts per billion by volume (ppbv) for methane at marine surface measurements sites \cite{NOAA} (an increase of over 1,000 ppbv in the past 250 years as a result of human activity \cite{IPCC2021}), and $\approx$413 parts per million by volume (ppmv) for carbon dioxide \cite{NOAA_CO2}; the mass of carbon dioxide in the atmosphere now is roughly 600 times the mass of methane. However, methane is a dramatically stronger absorber of thermal radiation than carbon dioxide. Despite its much lower concentration compared to carbon dioxide, methane is still responsible for trapping more than 50\% of the additional heat that atmospheric carbon dioxide traps compared to the pre-industrial era \cite{Balcombe2018}. \\

Satellite-borne remote sensing and monitoring of greenhouse gas emissions is playing an increasingly important role in assessing mankind’s impact on the climate \cite{Jacob2016,Palmer2020}, as direct measurements can displace or complement bottom-up inventory estimates that rely on self-reported industrial metrics \cite{OGMP2}. Measurements of methane column concentrations from space began in 2003 with the SCanning Imaging Absorption SpectroMeter for Atmospheric CHartographY (SCIAMACHY) on board ENVISAT, an ESA mission which terminated in 2012 \cite{Bovensmann1999}. SCIAMACHY was succeeded by the Greenhouse Gases Observing Satellite (GOSAT, 2009-present) and GOSAT2 (2018-present) \cite{Kuze2009,Glumb2014}, operated by JAXA. Both GOSAT and GOSAT2 have improved capabilities with pixel resolutions of 10x10 km$^2$ and global coverage in 3 days when compared to SCIAMACHY, which required 6 days for global coverage with a ground pixel resolution of 30x60 km$^2$. In 2016 the private enterprise  GHGSat-D instrument was launched, with a greatly improved pixel resolution over GOSAT of 50x50 m$^2$ for targeted viewing of selected methane sources \cite{Jacob2016}. Next generation instruments such as the Advanced Hyperspectral Imager (AHSI) launched on board China's GaoFen5 satellite in 2018 provide methane retrievals with a pixel resolution down to 30 metres \cite{AHSI2019}, but such missions are sporadically operated for targeted areas and do not provide the same level of spatial coverage as GOSAT \cite{Itziar2021}. Leading the field in regional observation is the TROPOspheric Monitoring Instrument (TROPOMI) on board the ESA's Sentinal-5 Precursor satellite. Launched in 2017, TROPOMI delivers daily global coverage, initially observing full methane columns with a ground pixel resolution of 7x7 km$^2$ at nadir \cite{Veefkind2012,Butz2011}, and later with an increased ground pixel resolution of 5.5x7 km$^2$ from August 2019 onwards. With early data successfully compared with GOSAT \cite{Hu2018}, TROPOMI is the cornerstone of the ESA's commitment to monitoring national pledges towards emission reductions under the Paris Climate Accord \cite{Clery2019}. \\

\begin{figure}[t]
\begin{center}
\makebox[0pt]{
\begin{minipage}{18.288cm}%
\subfloat{%
\includegraphics[width=18.288cm,keepaspectratio]{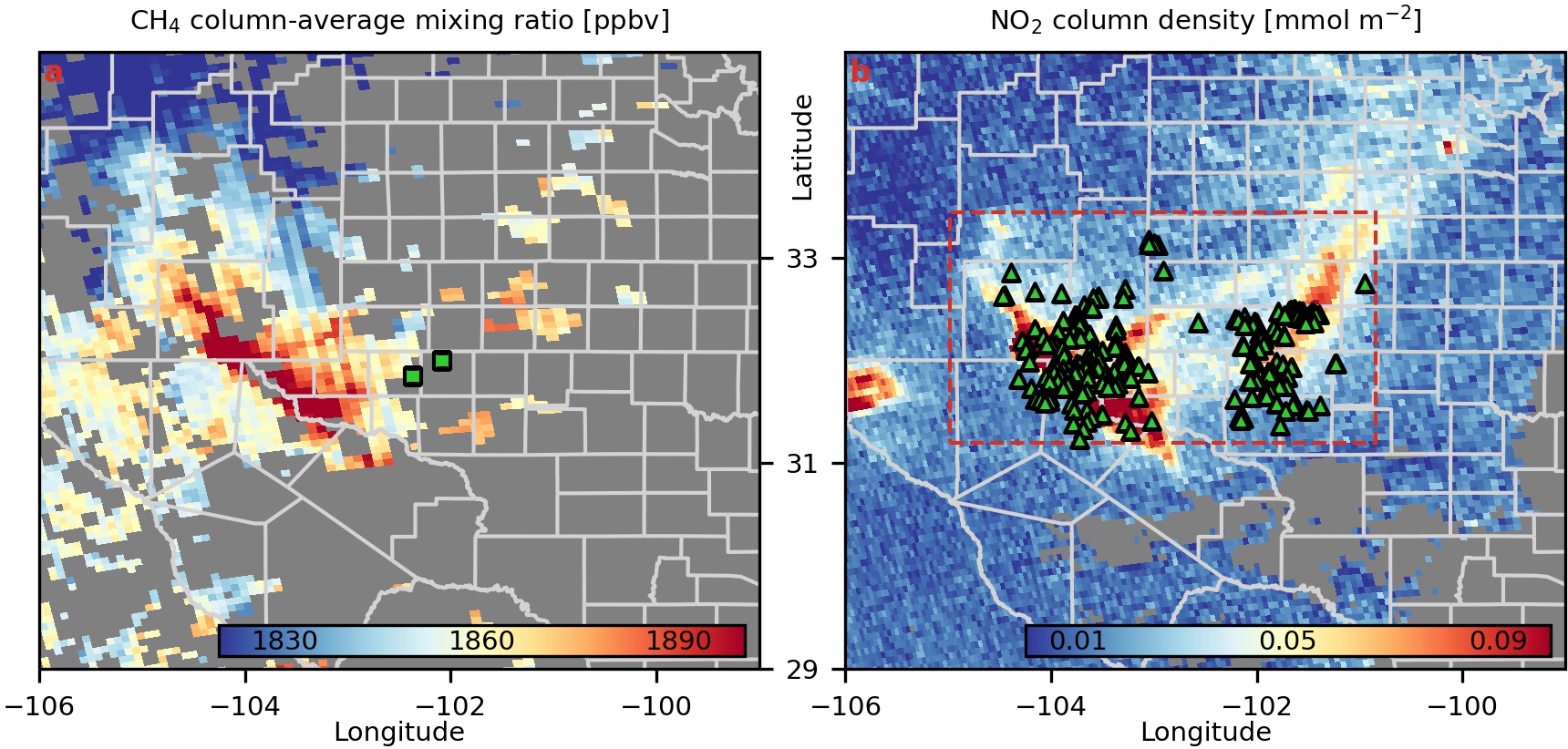}}%
\caption{TROPOMI observations of methane (\textbf{a}) and nitrogen dioxide (\textbf{b}) over the Permian Basin on January 31st, 2019. Only methane and nitrogen dioxide pixels which pass the recommended quality assurance threshold are shown \cite{CH4_ATBD} and all other pixels are masked. The study region is shown by the red dashed rectangle in \textbf{b}, superimposed over county lines which are shown in grey solid lines in both panels. We show the locations of the cities of Midland and Odessa (the only cities in the study region with populations exceeding 100,000 people) in \textbf{a} with green squares. Also shown in \textbf{b} with green triangles are the locations of lit methane flare stacks identified by VIIRS Nightfire \cite{Elvidge2013}, demonstrating their co-location with atmospheric overabundances of methane and nitrogen dioxide.}\label{fig:1}
\end{minipage}
}
\end{center}
\end{figure}

Remote sensing of methane emissions using satellites is attractive because it can be frequent, periodic, and non-intrusive to operations on the ground. Satellites therefore have the potential to detect intermittent methane sources or emissions released during abnormal operating conditions \cite{Alvarez2018}. Observations from SCIAMACHY \cite{Schneising2014} and GOSAT \cite{Turner2016} have successfully provided top-down estimates of regional methane emission rates. More recently, observations from TROPOMI have provided top-down estimates of methane emission rates from the Permian basin (shown in Fig. \ref{fig:1}, extending from western Texas to southeastern New Mexico) of 2.7 teragrams per year over a 12-month period from March 2018 to March 2019 \cite{Zhang2020}. \\

Although the TROPOMI instrument covers the entire surface of the Earth at least once a day, the actual amount of methane concentration data from TROPOMI is limited by a variety of factors \cite{Veefkind2012}. Cloud cover and the presence of aerosols in the atmosphere often hamper the retrieval of methane column data \cite{CH4_ATBD}. Consequently, sparse data is typically averaged over at least monthly timescales in order to obtain suitable coverage for emission estimates of oil and gas producing regions \cite{Schneising2014,Turner2016,Zhang2020}; however, individual daily TROPOMI methane observations can be used on an intermittent basis for flux estimates of smaller targets on clear-sky days \cite{Sadavarte2021}. In contrast, whilst TROPOMI observations of nitrogen dioxide are subject to similar difficulties posed by clouds and aerosols \cite{NO2_ATBD}, their geospatial coverage tends to exceed that of TROPOMI methane observations. Large excesses of nitrogen dioxide are often linked to regions of rapid urban expansion and industrial activity \cite{Duncan2016}. \\

Oil and gas production is also associated with nitrogen dioxide emissions. For example, lit methane flare stacks emit combusted natural gas which contains nitrogen dioxide \cite{Elvidge2018,Elvidge2016}; it is also known that the combustion efficiency in flares is not equal to 100\%, leading to the co-emission of methane into the atmosphere \cite{Zhang2020,Gouw2020}. Similarly, gas-fuelled compressors emit nitrogen dioxide during operation but they may also emit methane leaking through their seals. A final example are the pumps and storage tanks which are an integral part of oil distribution networks: pumps can emit nitrogen dioxide as a result of fuel combustion, whereas methane can leak from thief hatches in storage tanks. Indeed, a recent study \cite{Gouw2020} has shown that there is a correlation between the nitrogen dioxide and methane concentrations measured by TROPOMI over the Permian basin. \\

In this work, we develop a method to compensate for the missing direct methane data from TROPOMI by using nitrogen dioxide as a proxy of methane column density, with the methane-nitrogen dioxide relationship empirically inferred from sample locations where confident measurements exist for both species \cite{Gouw2020}. We develop a Bayesian model to infer missing methane column data based on co-located column values for nitrogen dioxide, which expands the spatial coverage of methane observations from TROPOMI. We use the Permian basin as a case study; as the most productive basin in the United States, it produces more than 18,000 million cubic feet of natural gas and nearly 5 million barrels of oil per day \cite{Permian2021}. We fit our model to TROPOMI observations, test its predictive ability and investigate how the inclusion of such estimates expands the daily spatial coverage of methane data. Spatial coverage over the Permian for the year 2019 is increased on average by 72\% when our model predictions are used to augment TROPOMI methane observations. We examine the implications of including inferred values in emission budgets by calculating the above-background mass of methane observed in the Permian basin over the course of the year 2019, and find the inclusion of inferred methane values results on average in nearly four more kilotonnes of excess methane observed per day. The spatial augmentation of TROPOMI methane observations will support inverse methods for estimating methane emission rates on shorter timescales than currently possible, which will be invaluable as policymakers begin to require recent and up-to-date methane emission estimates for industrial regions.

\section{Results}\label{sec:results}

\subsection{Model evaluation}
TROPOMI provides one observation of nitrogen dioxide and methane per day per location. Our model is a linear hierarchical model where the relationship between observations of nitrogen dioxide and methane on a given day $t$ are related to one another by a linear form with slope parameter $\beta_t$ and intercept parameter $\alpha_t$, plus another daily parameter to account for intrinsic scatter around the linear trend. We fit our model to a year's observations of methane and nitrogen dioxide over the Permian basin from 2019, using co-located observations contained within a marked study region shown in Fig. \ref{fig:1}. The daily parameter $\alpha_t$ has units of ppbv and roughly corresponds to the abundance of methane in the study region on day $t$ that is not associated with nitrogen dioxide. The daily parameter $\beta_t$ has units of ppbv / (mmol m$^{-2}$) as we scale all observations of nitrogen dioxide columns into mmol m$^{-2}$.\\

We specify a multivariate normal hierarchical prior on values of $\alpha_t$ and $\beta_t$ such that

\begin{align}
    \begin{pmatrix}
        \alpha_t \\
        \beta_t
    \end{pmatrix} &\sim \mathcal{N}\left(
    \begin{pmatrix}
        \mu_\alpha \\
        \mu_\beta
    \end{pmatrix},\;
    \begin{pmatrix}
        \Sigma_{11} & \Sigma_{12} \\
        \Sigma_{21} & \Sigma_{22}
    \end{pmatrix}
    \right) \label{eq:prior}
\end{align}

\noindent where $\Sigma_{11} = \sigma_\alpha^2$, $\Sigma_{22} = \sigma_\beta^2$ and $\Sigma_{12} = \Sigma_{21} = \rho\;\sigma_\alpha\;\sigma_\beta$. The hyperparameter $\rho$ controls the degree of correlation between $\alpha_t$ and $\beta_t$, with $\rho \in [-1,\;1]$. The hyperparameter $\mu_\alpha$ roughly corresponds to the average background level of methane in the study region across the year 2019, and the hyperparameter $\sigma_\alpha$ is the deviation that controls the spread of $\alpha_t$ around $\mu_\alpha$, with $\sigma_\alpha > 0$, and similarly for $\mu_\beta$ and $\sigma_\beta$. \\
  
We fit two models, each to a specific subset of data. First, we fit a ``data-rich" model to observations from days that produce numerous highly correlated co-located observations of methane and nitrogen dioxide in the study region. Afterwards, we fit a ``data-poor" model to observations from all other days where there are at least two co-located observations of methane and nitrogen dioxide in the study region. Note that the latter model can be fit to data very well and is referred to as the ``data-poor" model because it is fit to data from days that have relatively few co-located TROPOMI observations of methane and nitrogen dioxide. The difference in construction between the two models is that in the data-rich model, we stipulate a uniform prior independently for each of the hyperparameters in order for the model to learn their posteriors predominantly from the data, and in the data-poor model we provide the information learned from the data-rich model to the hyperparameters as prior information. We do this so that when the data-poor model is being fit either to sparse or less-correlated observations we are making full use of the information that can be obtained from data-rich days. The effect of requiring that data-rich days be those with both large amounts of co-located observations of nitrogen dioxide and methane as well as a high degree of correlation is that model predictions of methane from the fitted data-poor model may be more dominated by the prior when observations are extremely sparse, but we find that this is typically not the case. When fitting all models, we monitor the effective sample size (ESS) and $\hat{R}$ of all parameters in addition to the energy Bayesian fraction of missing information (E-BFMI) to ensure convergence and efficient sampling \cite{Betancourt2016,Vehtari2021}. The data-rich model was fit with an ESS of 212.9 and the data-poor model was fit with an ESS of 860.0, indicative of sufficient mixing in both cases. \\ 
 
\begin{figure}[t]
\begin{center}
\makebox[0pt]{\begin{minipage}{18.288cm}%
\subfloat{%
\includegraphics[width=18.288cm,keepaspectratio]{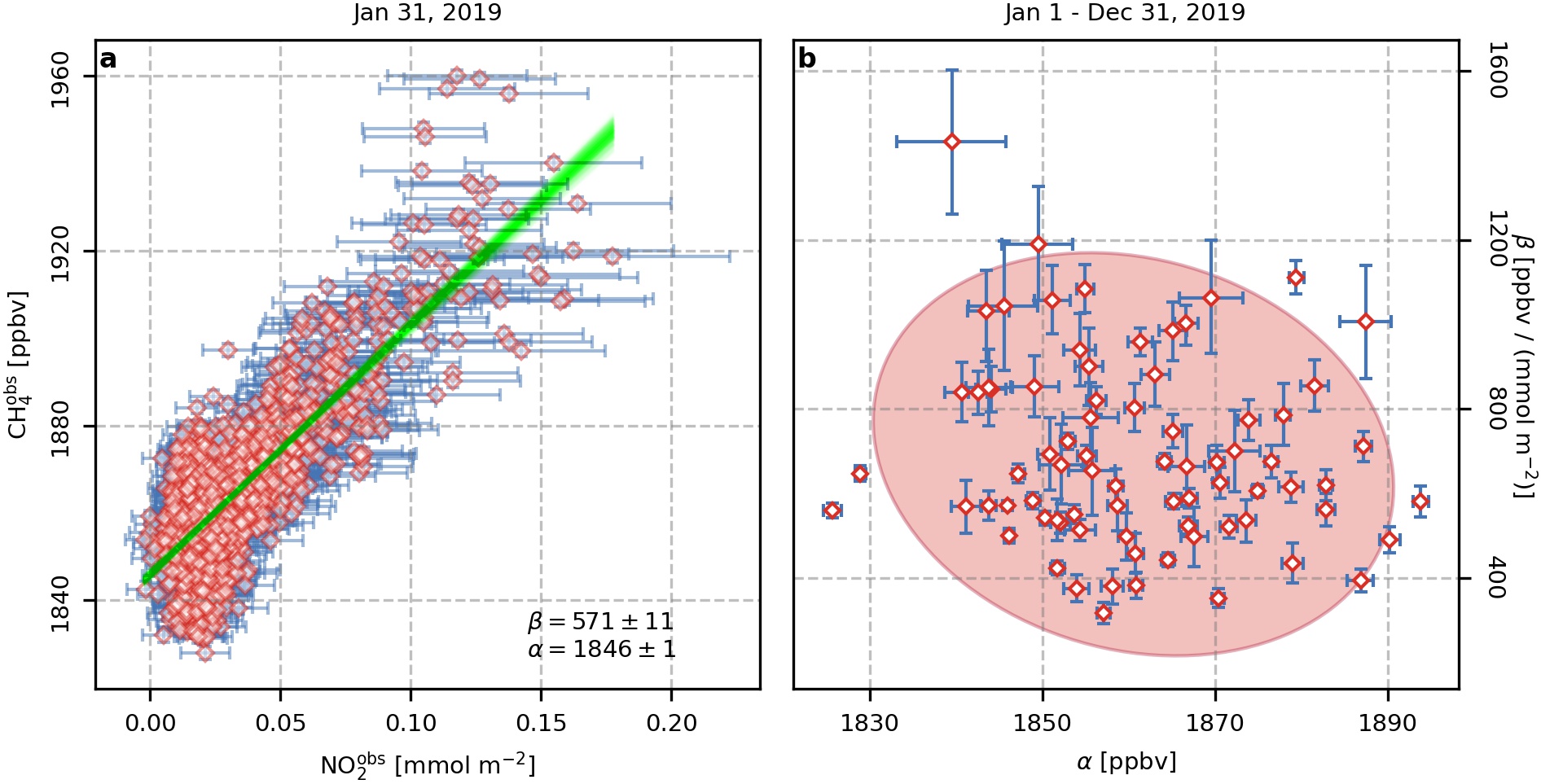}}
\caption{A graphical representation of how $\alpha_t$ and $\beta_t$ are estimated (\textbf{a}), with a demonstration of results accross all data-rich days in 2019 (\textbf{b}). \textbf{a} Co-located TROPOMI observations of methane column average dry air mixing ratio plotted against nitrogen dioxide column densities within the study region on January 31st, 2019, shown in red. Error bars on the observations (shown in blue) are the single-sigma precisions provided in the TROPOMI Level 2 Data Products. Superimposed over the scatterplot of observations we plot a random 1,000 selections of $(\alpha_t,\;\beta_t)$ pairs from the 4,000 sampled draws from the posterior of the fitted data-rich model, shown in lime green. In the bottom right corner we show median estimated posterior values of $\alpha_t$ [ppbv] and $\beta_t$ [ppbv / (mmol m$^{-2}$)] with $\pm$ indicating the extent of the 68\% central credible interval. \textbf{b} Median estimated posterior values of $\alpha_t$ and $\beta_t$ (shown in red) across all data-rich days with their estimated 68\% central credible intervals (shown in blue). Shown underneath the plotted values of $\alpha_t$ and $\beta_t$ is a red ellipse constructed from the median over random samples drawn from the joint posterior chain of $\mu_\alpha$, $\mu_\beta$, $\rho$, $\sigma_\alpha$ and $\sigma_\beta$. This ellipse indicates the information that is eventually supplied to the data-poor model as prior information for $\alpha_t$ and $\beta_t$.}\label{fig:2}
\end{minipage}}
\end{center}
\end{figure}

We examine the joint posterior distribution of the hyperparameters of the data-rich model, and represent median values of $\mu_\alpha$, $\mu_\beta$, $\rho$, $\sigma_\alpha$ and $\sigma_\beta$ in panel \textbf{b} of Fig. \ref{fig:2}. We find that the correlation $\rho$ between $\alpha_t$ and $\beta_t$ on data-rich days for the year 2019 has a median value of -0.18 with a 68\% central credible interval of (-0.30, -0.05). We would expect a value of $\rho < 0$ due to the positive correlation between the amount of methane and nitrogen introduced to the atmosphere from flare stack combustion chemistry \cite{Sonibare2004,Deetz2017,Ismail2016,Umukoro2017}, i.e., positive correlation in the data implies negative correlation between the slope and intercept. However, it is important to note that this result was obtained without stipulating any prior constraint that $\rho$ must be negative. Multivariate normal hierarchical prior specification and uniform hyperpriors for hyperparameters in the data-rich model allows us to learn the degree to which the daily model parameters $\alpha_t$ and $\beta_t$ are correlated, and leverage this information as a prior of appropriate strength in the data-poor model. \\

Other model hyperparameters that we discuss at this stage are $\mu_\alpha$ and $\sigma_\alpha$. These two model parameters have physical meaning, where $\mu_\alpha$ can be thought of as the `mean' regional background of $\mathrm{CH}_4$ in our study region for the year 2019, and $\sigma_\alpha$ as the extent to which the daily intercept $\alpha$ varies around $\mu_\alpha$ from day to day. We find $\mu_\alpha$ to have a median value of 1861.32 ppbv with a 68\% central credible interval of (1859.52, 1863.15), and $\sigma_\alpha$ to have a median value of 14.44 ppbv with a 68\% central credible interval of (13.25, 15.73). \\

\subsection{Predictive skill}\label{sec:predictive_skill}
\begin{figure}[t]
\begin{center}
\makebox[0pt]{\begin{minipage}{18.288cm}%
\subfloat{%
\includegraphics[width=18.288cm,keepaspectratio]{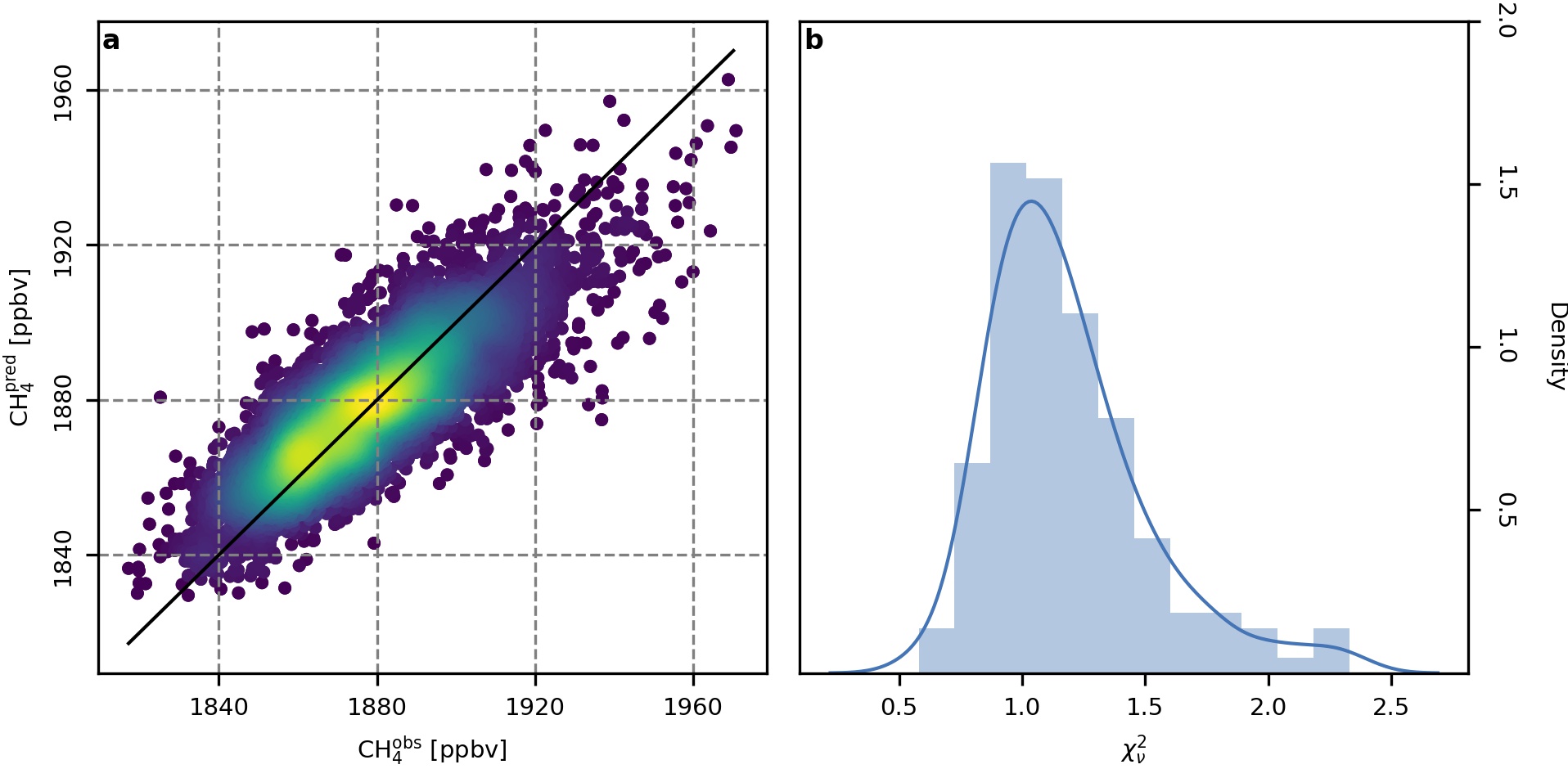}}
\caption{Results of the dropout testing. \textbf{a} Predictions of methane $\mathrm{CH}_4^{\mathrm{pred}}$ plotted against actual TROPOMI observations $\mathrm{CH}_4^{\mathrm{obs}}$. The color scale indicates the number density of the scatterplot. Predictions come from a model that was fitted without using the observed values shown on the x-axis. We plot in black the line $y=x$ to demonstrate the effect that uncertainty on nitrogen dioxide observation has on our model; regression dilution results in slight underestimation of methane at high values of nitrogen dioxide and overestimation at low values of nitrogen dioxide, though there is still good agreement overall. \textbf{b} A distribution of reduced chi-squared values for data-rich days in the year 2019, comparing model predictions to the data withheld from the model fitting.}\label{fig:dropout_validation}
\end{minipage}}
\end{center}
\end{figure}

To assess the predictive ability of our models, we performed hold-out testing by refitting our models to a random selection of 80\% of observations in the study region on each day, using the remaining 20\% of TROPOMI methane observations for comparison against model predictions at those locations. Both the data-rich and the data-poor model fit to the 80\% datasets with satisfactory E-BFMI, ESS and $\hat{R}$. After generating predictions from the fitted model and comparing to co-located withheld observations across all days, we calculate values of reduced chi-squared $\chi_{\nu}^2$ for each day (i.e., chi-squared per degree of freedom), the resulting distribution of which is shown in panel \textbf{b} of Fig. \ref{fig:dropout_validation}. We also calculate residuals between model predictions and withheld observations across all days, which have a mean of 0.15 ppbv and standard deviation 12.09 ppbv, correlated with Pearson $R=0.82$ (shown in panel \textbf{a} of Fig. \ref{fig:dropout_validation}). For comparison, first results from TROPOMI were initially tested against co-located GOSAT observations with a mean difference of 13.6 ppbv and standard deviation 19.6 ppbv, correlated with Pearson $R=0.95$ \cite{Hu2018}. Recent work has derived the observation standard deviation over the Permian to be 11 ppbv \cite{Zhang2020}. In an extreme case where all 11 ppbv is independent of our model error, the standard error of our model predictions would be roughly $\sqrt{12^2 + 11^2} \approx 16$ ppbv. Panel \textbf{a} of Fig. \ref{fig:dropout_validation} also demonstrates the tendency of the model to underestimate methane from high values of observed nitrogen dioxide and overestimate methane at low values of nitrogen dioxide. This is a result of regression dilution caused by the relatively high uncertainty of the TROPOMI observations of nitrogen dioxide \cite{Hutcheon2010,Andreon2013}, seen in panel \textbf{a} of Fig. \ref{fig:2}. Correcting for regression dilution is not necessary in predictive modelling scenarios. Validation studies have shown that the TROPOMI nitrogen dioxide data product can be biased low by as much as 50\% over highly polluted regions when compared to ground-based observations \cite{Verhoelst2021}. Since we fit our models using nitrogen dioxide TROPOMI observations that are not bias corrected, the resulting predicted methane abundances will not be degraded when non-bias-corrected TROPOMI nitrogen dioxide observations are used as input for the methane predictions.  \\

\begin{figure}[t]
\begin{center}
\makebox[0pt]{\begin{minipage}{18.288cm}%
\subfloat{%
\includegraphics[width=18.288cm,keepaspectratio]{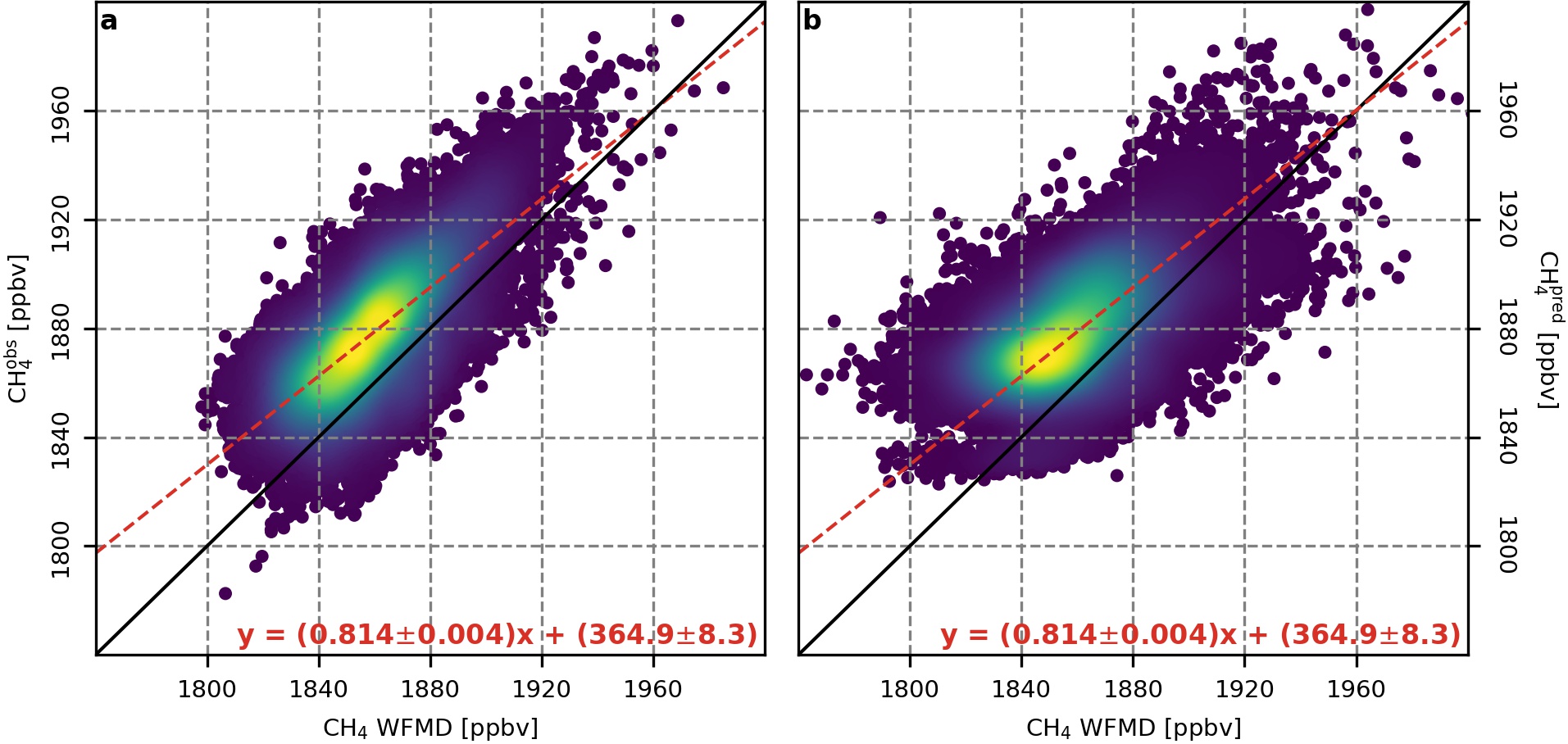}}
\caption{\textbf{a} A comparison of TROPOMI observations of methane CH$_4^{\mathrm{obs}}$ over the Permian in the year 2019 on data-rich days to co-located observations from the WFMD CH$_4$ data product \cite{WFMD2019}. In black we plot the line $y=x$ and with the red dashed line we plot the ordinary least squares line of best fit (the parameters of the fit are shown in the bottom right of the panel). The color scale indicates the number density of the scatterplot. The root mean square deviation of the data around the line of best fit is 15.5 ppbv. CH$_4^{\mathrm{obs}}$ correlates with CH$_4$ WFMD with a Pearson R of 0.73. \textbf{b} a comparison of methane predictions CH$_4^{\mathrm{pred}}$ from our fitted model over the Permian in the year 2019 on data-rich days to co-located observations from the WFMD CH$_4$ data product. As in \textbf{a}, the color scale indicates the number density of the data, and in black we plot the line $y=x$. We plot dashed in red the same line from \textbf{a}, not a fit to the data in \textbf{b}. The root mean square deviation of the data in \textbf{b} around this same regression line is 15.8 ppbv. CH$_4^{\mathrm{pred}}$ correlates with CH$_4$ WFMD with a Pearson R of 0.58 .}\label{fig:bremen_correlation}
\end{minipage}}
\end{center}
\end{figure}

We also assess the predictive ability of our models against the Weighting Function Modified Differential Optical Absorption Spectroscopy (WFMD) CH$_4$ data product \cite{WFMD2019}. This data product has much better spatial coverage than the operational Sentinal-5P CH$_4$ data product that we fit our models to. We find that predictions of methane from our fitted model correlate nearly as well with co-located observations of the Permian basin from the WFMD CH$_4$ data product as the ``original" TROPOMI methane observations (Fig. \ref{fig:bremen_correlation}). TROPOMI observations of methane exhibit a standard deviation of 15.5 ppbv around their line of best fit with co-located observations from the WFMD CH$_4$ data product red with a correlation of Pearson $R=0.73$, whilst predictions from our model exhibit a standard deviation of 15.8 ppbv around the same line red with a Pearson $R=0.58$.

\subsection{Seasonality}
\begin{figure}[t]
\begin{center}
\makebox[0pt]{\begin{minipage}{18.288cm}%
\subfloat{%
\includegraphics[width=18.288cm,keepaspectratio]{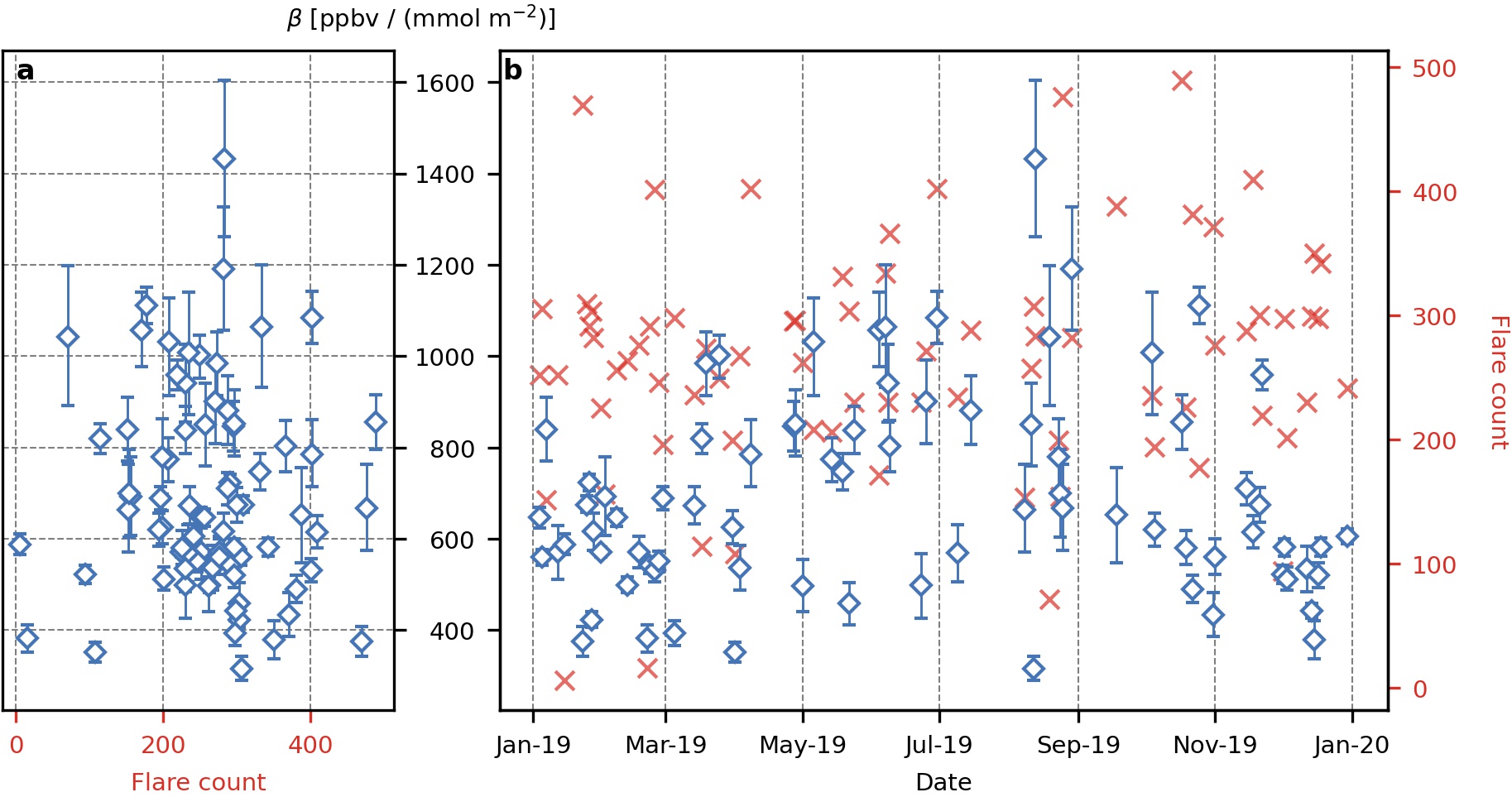}}
\caption{A demonstration that the number of total lit methane flares in the study region identified by VIIRS does not appear to correlate with estimated values of $\beta_t$. In panel \textbf{a} we plot median estimated posterior values of $\beta_t$ (with 68\% central credible regions as error bars) as a function of the identified number of lit methane flares on that day. \textbf{b} The same estimates of $\beta_t$ from panel \textbf{a} as a time series. Also shown with red x's the numbers of active lit methane flares identified by VIIRS Nightfire on data-rich dates. In general, higher values of $\beta_t$ are found to occur in the summer months. This could be due to the fact that nitrogen dioxide is removed from the atmosphere more quickly in the summer than in the colder winter months, or due to changes in operating procedure when energy demand is low.}\label{fig:3}
\end{minipage}}
\end{center}
\end{figure}

\begin{figure}[t]
\begin{center}
\makebox[0pt]{\begin{minipage}{18.288cm}%
\subfloat{%
\includegraphics[width=18.288cm,keepaspectratio]{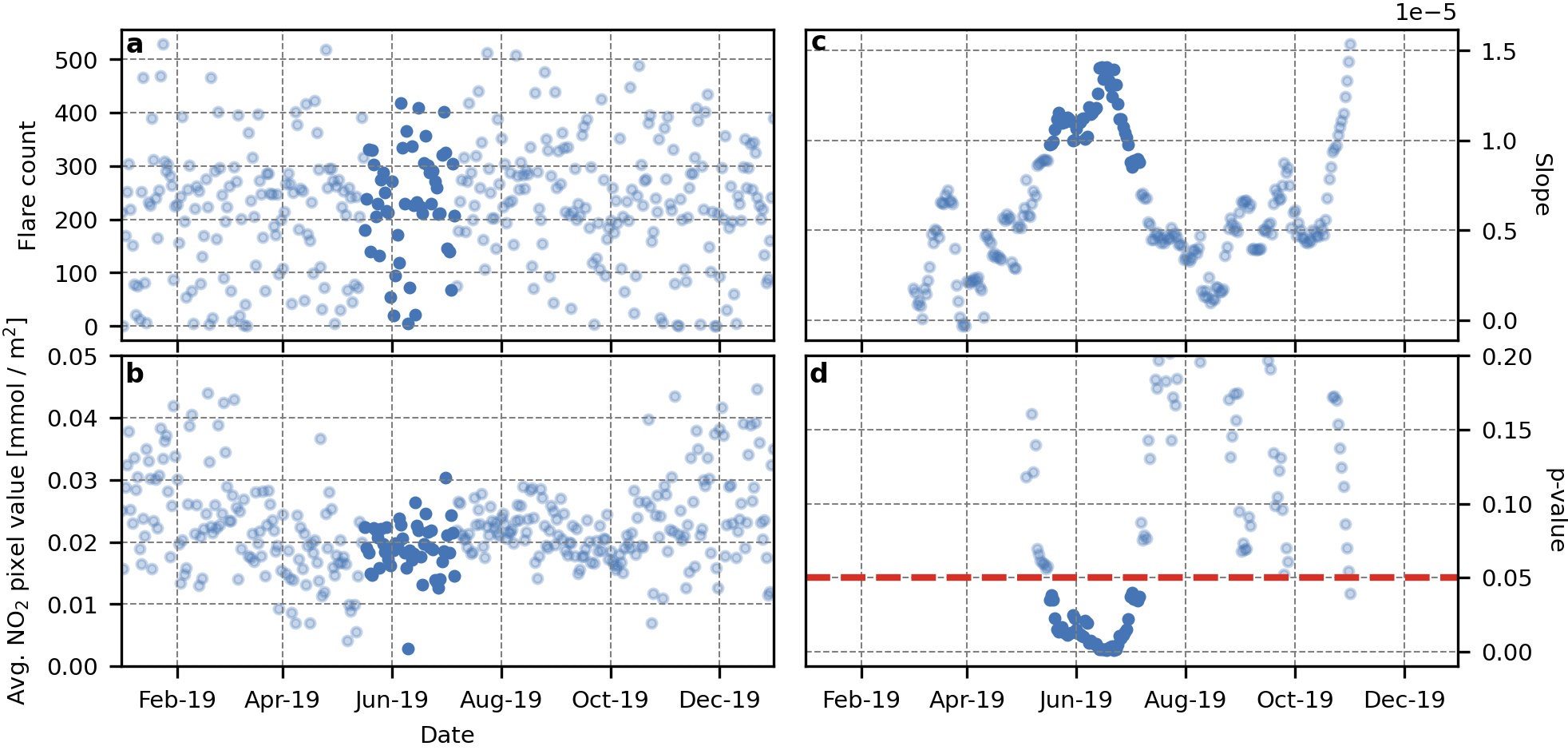}}
\caption{\textbf{a} A time series of active flares identified by VIIRS Nightfire in the study region in the year 2019. \textbf{b} a time series of average TROPOMI nitrogen dioxide pixel value in the study region. \textbf{c} A time series of ordinary least squares slope estimates, fitted to average nitrogen dioxide pixel value against flare count, using rolling 120-day averages. \textbf{d} p-values for the slopes calculated in \textbf{b} (values exceeding 0.20 emitted). In all panels, only data on dates where the slopes in \textbf{c} are significantly positive according to \textbf{d} are shown as solid blue, demonstrating that nitrogen dioxide and number of identified flares are strongly correlated in the summer months.} \label{fig:no2_flare_stack_correlation}
\end{minipage}}
\end{center}
\end{figure}

After re-fitting the data-rich model to 100\% of available observations we examine posterior estimates of daily model parameters as time series. We plot median values of the slope parameter $\beta_t$ in Fig. \ref{fig:3} along with a time series of active flare stacks in the study region on data-rich days, identified from VIIRS Nightfire \cite{Elvidge2013}. We do not find any significant correlation between estimated $\beta_t$ and the total number of identified flare stacks inside the study region on a given day, which is not unexpected as recent work suggests that super-emitting individual flares may be accountable for the majority of methane emissions in the Permian basin \cite{Itziar2021}. However, we do find that higher values of $\beta_t$ tend to occur in the summer months. We investigate this further by examining the seasonality of nitrogen dioxide in the study region, shown in of Fig. \ref{fig:no2_flare_stack_correlation} \textbf{c}. Nitrogen dioxide has a shorter atmospheric lifetime in the warmer summer months, which could explain the higher inferred values of $\beta_t$ shown in Fig. \ref{fig:3}. However, in Fig. \ref{fig:no2_flare_stack_correlation} we also investigate the correlation between the average value of nitrogen dioxide in the study region and number of active flares, and find that nitrogen dioxide correlates with number of flares to a significant level in the warmer summer months. Future work is needed to determine the origin of this correlation.

\subsection{Enhancement of spatial coverage}
\begin{figure}
\begin{center}
\makebox[0pt]{\begin{minipage}{18.288cm}%
\subfloat{%
\includegraphics[width=18.288cm,keepaspectratio]{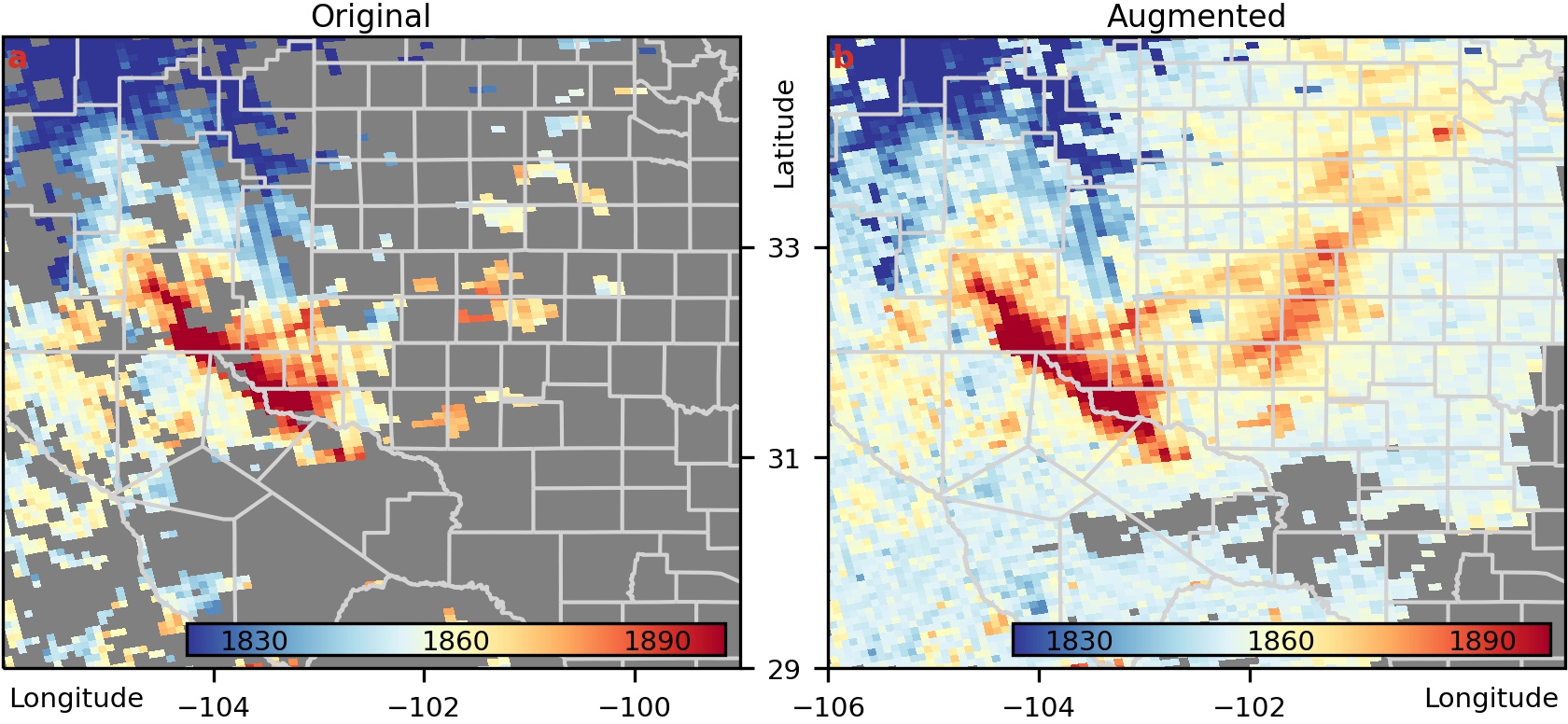}}
\caption{A comparison between original (\textbf{a}) and augmented (\textbf{b}) TROPOMI observations of methane columns. The colorbar labels in both plots are in units of ppbv. \textbf{a}  The same TROPOMI observation of methane columns as in Fig. \ref{fig:1} \textbf{a}. \textbf{b} The same observations from \textbf{a}, augmented with predictions from the fitted model at all locations where we have a TROPOMI observation of nitrogen dioxide that passes the recommended quality assurance thresholds. On this day, including model predictions increased the spatial coverage of the study region from 53\% to 100\%. } \label{fig:4}
\end{minipage}}
\end{center}
\end{figure}

\begin{figure}
\begin{center}
\makebox[0pt]{\begin{minipage}{18.288cm}%
\subfloat{%
\includegraphics[width=18.288cm,keepaspectratio]{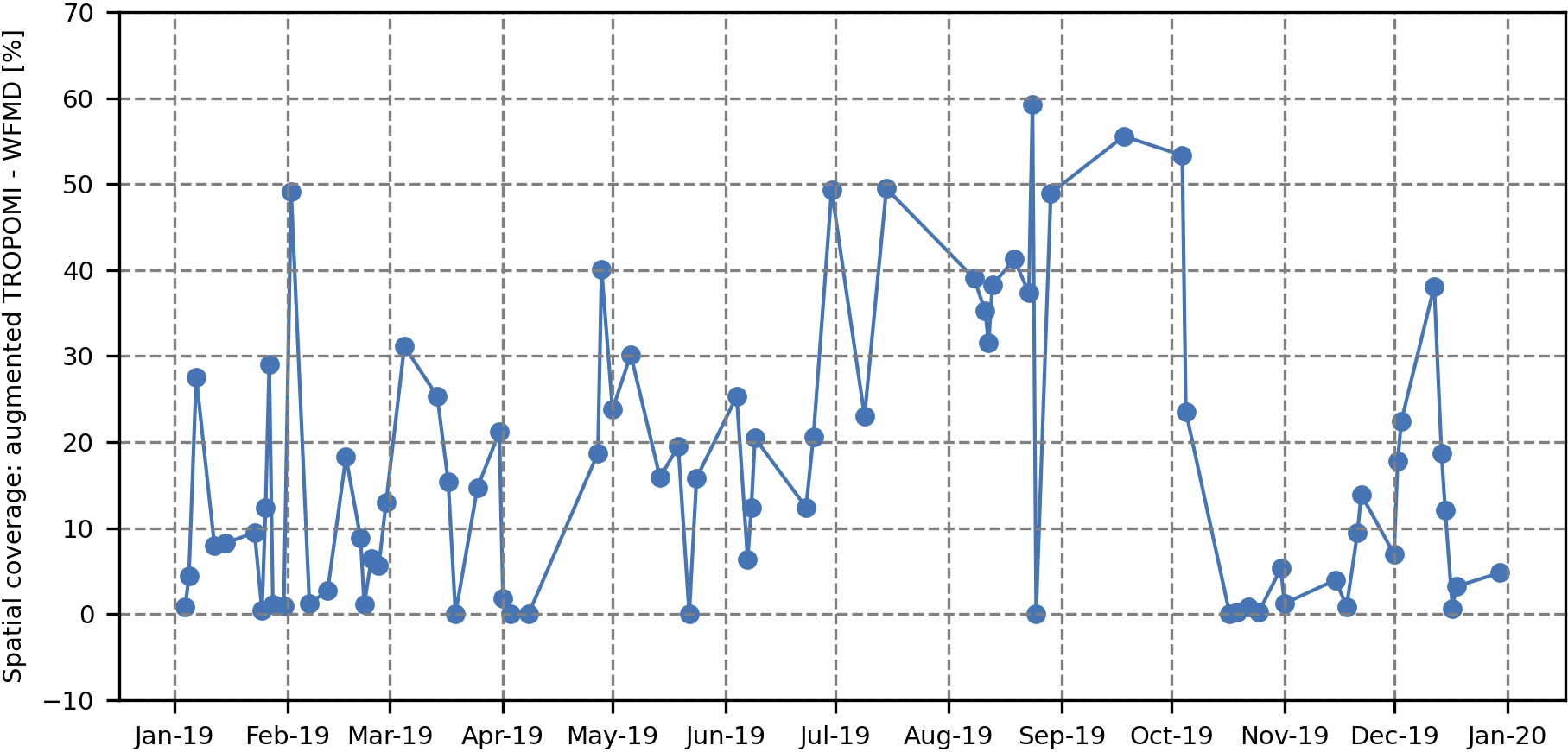}}
\caption{A time series of the difference between the spatial coverage of our augmented TROPOMI methane observations and the spatial coverage of the WFMD CH$_4$ data product. In general, the spatial coverage of our augmented TROPOMI observations over the Permian Basin in the year 2019 exceeds that of the WFMD CH$_4$ data product.} \label{fig:bremen_spatial_coverage}
\end{minipage}}
\end{center}
\end{figure}

We augment daily observations over the study region by including model predictions at locations where direct observations are not available from TROPOMI. Predictions are only made from co-located TROPOMI observations of nitrogen dioxide with an associated quality assurance value greater than or equal to 0.75. An example of this procedure is shown in Fig. \ref{fig:4}. Prior to augmentation using predictions from the model, we calculate that the mean value of daily pixel coverage within the study region for the year 2019 is $16\%\pm13\%$. After adding in model predictions at appropriate pixel locations, the mean value of daily pixel coverage rises to $88\%\pm18\%$. A time series demonstrating this rise in spatial coverage is shown in Fig. \ref{fig:5} \textbf{a}. We find that the spatial coverage of our augmented TROPOMI observations at least matches and usually exceeds that of the WFMD CH$_4$ data product (Fig. \ref{fig:bremen_spatial_coverage}). Increasing the spatial coverage over regions of interest like the Permian basin can allow for the aggregation of data on shorter timescales than previously used in performing methane emission estimates, which will be key for accurately monitoring greenhouse gas emissions in near-real time.

\subsection{Examination of urban influence}
The cities of Odessa and Midland are contained within the study region and each have populations that exceed 100,000 people. Cities are known to emit large quantities of nitrogen dioxide \cite{Duncan2016}, and may emit co-located methane at a rate that differs from the rural surrounding oil and gas producing regions. We here examine the extent to which including these cities in the study region may affect predictions of methane in non urban areas. To do this, we define a new sub-region within the study region that contains the two cities. Fig. \ref{fig:city_influence} shows comparative distributions of various TROPOMI observations and model predictions, segregated according to whether or not they are located in the urban sub-region containing the cities. Using a Kolmogorov-Smirnov test, we determine these pairs of distributions to be significantly different from one another \cite{KSTest1951}. We calculate the fractional difference between the mean value of observed nitrogen dioxide columns over the urban sub-region and the mean value of observed nitrogen dioxide columns over the surrounding rural areas and find that they differ by 31.28\%. This indicates that nitrogen dioxide emissions tend to be higher on average over the city sub-region than over the surrounding rural regions, and that as a result it may be the case that methane emissions are slightly over-predicted by our model over the city sub-region. However, the difference between the mean values of predicted and observed methane over the urban sub-region is less than one percent. The same is true for the fractional difference between the mean values of predictions and observations of methane over the rural surroundings. Although it is beyond the scope of this work to fully investigate this result, it may be the case that our model under-predicts methane emissions in the surrounding rural areas of the study region where methane sources like livestock are not directly correlated with nitrogen dioxide emissions. We note that the majority of locations of data in our sample correspond to rural areas, and Sec. \ref{sec:predictive_skill} demonstrated that in general our model predictions agree well with the data. In future work, this methodology could be applied to smaller study areas where infrastructure remains more homogenous, though this would come at a cost of less data.

\subsection{Observed mass of methane hotspots}
By subtracting a nominal background level of methane from TROPOMI observations and model predictions, we can calculate the above-background mass of methane contained above the study region on a given day. We use the global monthly marine mean surface value of methane as a far-field reference background \cite{NOAA}.\\

We convert all methane columns within the study region on each day to above-background masses and integrate over the spatial extent to calculate the above-background mass. This involves converting all values of methane (which either returned from TROPOMI or predicted from the model as dry-air column averaged mixing ratios) into column densities, for which we need a grid of dry-air column densities at all pixel locations. We calculate a grid of dry air column density to convert both TROPOMI observations and model predictions to masses in a consistent manner \cite{ERA5}, since dry air column densities are only provided with the TROPOMI Level 2 CH$_4$ Data Product at the location of observed pixels \cite{CH4_ATBD}. Values of dry air column density from our grid are correlated with values returned with the TROPOMI Level 2 $\mathrm{CH}_4$ Data Product with a Pearson R of 0.97, and have a mean residual of -1,242 kg m$^{-2}$ with standard deviation 1,836 kg m$^{-2}$. We use the root mean square deviation of 2,217 kg m$^{-2}$ for error propagation in the calculation of mass of methane contained with a pixel in the study region. \\

We find that over the course of the year 2019, the average daily above-background mass of methane is $0.9\pm1.5$ kilotonnes. We re-calculate this quantity including predictions from the model and find a new daily mean of above-background methane mass of $4.3\pm5.1$ kilotonnes. These two quantities are plotted in panel \textbf{c} of Fig. \ref{fig:5}. The choice of a reference background from the far field does allow for negative values of above-background methane levels, but the point of the calculation is to demonstrate the affect of including predictions, and so the choice of background isn't strictly important. Previous work has demonstrated that time-integrated TROPOMI observations of methane can be used to create top-down estimates of methane emission rates \cite{Zhang2020}. While we do not currently carry out any inverse analysis, we demonstrate that the inclusion of predictions of methane from co-located observations of nitrogen dioxide increases the above-background value of methane loading in the study region by kilotonnes per day on average, which could indicate that the inclusion of predictions reveals methane hotspots that were previously unobserved. By increasing the spatial coverage of observations, we can in future work attempt to increase the temporal resolution of methane emission rates in the Permian.

\section{Discussion}\label{sec:discussion}
The remote sensing of atmospheric methane from space is a crucial tool for monitoring anthropogenic greenhouse gas emissions. Using the Permian basin as a case study, we show that observations of nitrogen dioxide can be used as predictors for methane at locations for which meteorological factors prevent TROPOMI from making direct observations of methane. We validate our Bayesian model using a variety of metrics and examine its predictive ability against withheld observations. When using predicted values of methane observations to augment daily observations over the Permian, we find that methane estimates can be obtained with effectively full spatial coverage on most days, and that the observed mass of methane hotspots is increased by approximately 3.4 kilotonnes per day on average, a 377\% increase. \\

The algorithm described in this work has limitations that could be improved upon in future work. The atmospheric lifetimes of methane and nitrogen dioxide are different, with nitrogen dioxide being removed from the atmosphere on a scale of minutes compared to years for methane. This effect is enhanced in warmer summer months, when the atmospheric lifetime of nitrogen dioxide decreases even further compared to that of methane. It is therefore possible that our model may tend to underestimate methane emissions as a consequence of ``missing" nitrogen dioxide that has reacted away, leaving behind overabundances of methane that are no longer co-located with an overabundance of nitrogen dioxide. This effect might be incorporated as a smoothly varying seasonal $\beta_t$. Inclusion of wind field data may allow for the temporal modeling of nitrogen dioxide decay as pollutants move through plumes. Furthermore, varying surface altitudes may result in spurious methane predictions if the current model is used for a study region with highly varying topography. \\

Whilst our current methodology works efficiently for a study region like the Permian basin where oil and gas producing infrastructure results in highly correlated overabundances of nitrogen dioxide and methane, it remains to be seen if the methodology will be as easily applicable to other regions. Whereas our model does not require methane and nitrogen dioxide to be exactly co-emitted, it does require them to be roughly correlated on approximately 5 km scales. Ongoing work is investigating the applicability of the model to additional regions of varying geographies and industrial settings. \\

This work presents what we believe is thus far the most extensive estimation of atmospheric methane over the Permian basin from satellite data. The current methodology might provide a useful starting point for joint inversion of satellite observations of nitrogen dioxide and methane, and hence potentially estimation of methane emission rates on timescales much shorter than previously achieved.

\section{Data Availability}\label{sec:data_availability}
The data that support this study are available from the Copernicus Open Access Hub (for TROPOMI $\mathrm{CH}_4$ and $\mathrm{NO}_2$ Level 2 Products), the Copernicus Climate Change Service (for ERA5 reanalysis data), the Earth Observation Group at the Colorado School of Mines (for VIIRS Nightfire) and the Global Monitoring Laboratory at the National Oceanic and Atmospheric Association (for mean global marine surface methane). 

\section{Code Availability}\label{sec:code_availability}
All code used in this study to analyse data and generate figures can be found at the corresponding author's Github repository titled \textit{TROPOMI\_LHM} (``Linear Hierarchical Model"). 

\section{Acknowledgements}\label{sec:acknowledgements}
C.R. acknowledges financial support from Shell Research Ltd. through the Cambridge Centre for Doctoral Training in Data Intensive Science grant number ST/P006787/1. C.R. thanks the referees for their time and expertise in reviewing the manuscript.

\section{Author Contributions}\label{sec:author_contributions}
C.R. retrieved all data, wrote all code and conducted the analysis of this work. O.S. supervised the project and guided the analysis. K.M., P.J. and M.J. guided the construction of the Bayesian model. R.I. and B.H. contributed discussions of the data, satellites and industrial context.

\section{Competing Interests}\label{sec:competing_interests}
We report no competing interests.

\newpage

\printbibliography

\newpage

\section{Methods}\label{sec:methods}
For each day in 2019 we retrieved TROPOMI observations of methane and nitrogen dioxide over the study region as shown in Fig. \ref{fig:1}. Before conducting any analysis, we reduce the data by ignoring pixels that do not pass the recommended quality assurance (qa) value. For the TROPOMI Level 2 $\mathrm{CH}_4$ Data Product, the threshold qa factor for recommended usage is 0.5 or greater, and for the TROPOMI Level 2 $\mathrm{NO}_2$ Data Product, the threshold qa factor for recommended usage is 0.75 or greater. We then classified individual days as being either rich or poor in remaining data, with data-rich days being those with $N\geq100$ co-located observations of methane and nitrogen dioxide in the study region correlated with Pearson's $R\geq0.4$, and data-poor days being all other days with $N\geq2$. Co-located observations are determined by linearly interpolating the grid of $\mathrm{CH}_4^{\mathrm{obs}}$ onto the grid of $\mathrm{NO}_2^{\mathrm{obs}}$. We leave the TROPOMI Level 2 $\mathrm{CH}_4$ Data Product in units of ppbv, but we scale the TROPOMI Level 2 $\mathrm{NO}_2$ Data Product from mol m$^{-2}$ to mmol m$^{-2}$.

\subsection{Data-rich model}\label{sec:data-rich_model}
We next developed a fully Bayesian linear hierarchical model to fit solely to observations from data-rich days. Data-rich days are indexed via $t=1,2,...,D$, and co-located observations of methane and nitrogen dioxide within the study region on day $t$ are indexed via $i=1,2,...,N$. \\

\noindent We relate observed values of methane and nitrogen dioxide to their true latent values via 

\begin{align}
\mathrm{NO}_{2,i}^{\mathrm{obs}}
&\sim \mathcal{N}\left(\mathrm{NO}_{2,i}^{\mathrm{true}},\,\sigma_{\mathrm{N},i}^2\right) \label{eq:latent_no2}\\
\mathrm{CH}_{4,i}^{\mathrm{obs}} &\sim \mathcal{N}\left(\mathrm{CH}_{4,i}^{\mathrm{true}},\,\sigma_{\mathrm{C},i}^2\right)\label{eq:latent_ch4}
\end{align}

\noindent where $\sigma_{\mathrm{N}i}$ and $\sigma_{\mathrm{C}i}$ are the TROPOMI-provided measurement standard deviations on $\mathrm{NO}_{2i}^{\mathrm{obs}}$ and  $\mathrm{CH}_{4i}^{\mathrm{obs}}$ respectively. We also relate latent values of methane to latent values of nitrogen dioxide via 

\begin{equation}
    \mathrm{CH}_{4,i}^{\mathrm{true}} \sim \mathcal{N}\left(\alpha_t + \beta_t\,\mathrm{NO}_{2,i}^{\mathrm{true}},\,\gamma_t^2\right)\label{eq:linear}
\end{equation}

\noindent where we have now introduced the daily model parameters $\alpha_t$, $\beta_t$ and $\gamma_t$. On a given day $t$, $\alpha_t$ and $\beta_t$ are respectively the y-intercept and slope of the line of best fit relating methane to nitrogen dioxide, while $\gamma_t$ is the standard deviation of the scatter around the mean relation. \\

We stipulate an improper flat prior on latent values of nitrogen dioxide observations in order to combine equations \eqref{eq:latent_no2}, \eqref{eq:latent_ch4} and \eqref{eq:linear} into the single model equation 

\begin{equation}
    \mathrm{CH}_{4,i}^{\mathrm{obs}} \sim \mathcal{N}\left(\alpha_t + \beta_t\,\mathrm{NO}_{2,i}^{\mathrm{obs}},\,\beta_t^2\,\sigma_{\mathrm{N},i}^2 + \gamma_t^2 + \sigma_{\mathrm{C},i}^2\right)\label{eq:likelihood}.
\end{equation}

\noindent Writing this equation in this way is desirable because it relates the observed pixel value of methane to the observed pixel value of nitrogen dioxide entirely in terms of the associated pixel precisions and the by-day model parameters $\alpha_t$, $\beta_t$ and $\gamma_t$. \\

We add hyperparameters to our model by including a multivariate prior distribution for $\alpha_t$ and $\beta_t$, shown in equation \eqref{eq:prior}. We include this multivariate prior to allow for the possibilty of learning the extent to which $\alpha_t$ and $\beta_t$ are correlated. Although we believe it is likely that $\alpha_t$ and $\beta_t$ are negatively correlated, we do not encode this belief in the prior. We instead assume a uniform flat prior on the domain $\left(-\infty,\;\infty\right)$ on each of $\mu_\alpha$, $\mu_\beta$, $\Sigma_{12}$ and $\Sigma_{21}$, and assume a uniform flat prior on the domain $\left(0,\;\infty\right)$ on both $\Sigma_{11}$ and $\Sigma_{22}$. This lets us learn the extent to which $\alpha_t$ and $\beta_t$ are correlated entirely from the posterior inference. Equations \eqref{eq:likelihood} and \eqref{eq:prior} are the likelihood and prior of our data-rich model respectively. We write our model in Stan \cite{Stan} via the interface CmdStanPy \cite{CmdStanPy} and sample the posterior of our model using Stan's default Markov Chain Monte Carlo algorithm NUTS (the No U-Turn Sampler, which is a variant of Hamiltonian Monte Carlo). When fitting our model we specify the algorithm to draw samples from the posterior using four separate Markov chains, each with 500 burn-in iterations with a further 1000 retained draws, combining for a total of 4000 draws from the posterior for each of our model parameters. Specifying 1000 post burn-in draws per chain easily allows for a sufficient ESS.\\

\subsection{Data-poor model}\label{sec:data-poor_model}
We developed a separate Bayesian model to fit to days that we identified as being poor in data. Data-poor days were classified as those that were not data-rich and have $N\geq2$ co-located observations of methane and nitrogen dioxide in the study region. The point of developing a second model to fit to data-poor days after we have already fit a model to data-rich days is to incorporate information learned from the data-rich model into the data-poor model as an informed prior. \\

As in the data-rich model, the likelihood of the data-poor model is given by equation \eqref{eq:likelihood}, and the multivariate prior on values of $\alpha_t$ and $\beta_t$ is given by equation \eqref{eq:prior}. However, when fitting the hierarchical model to all data-rich days simultaneously, we've learned information about what sort of values the model hyperparameters ``should" take. To account for this information we've learned, we add a prior on the hyperparameters with 

\begin{align}
    \begin{pmatrix}\mu_\alpha\\
                   \mu_\beta \\
                   \Sigma_{11}\\
                   \Sigma_{12}\\
                   \Sigma_{22}\\
    \end{pmatrix} &\sim \mathcal{N}\left(\theta,\,\Upsilon\right).
\end{align}

\noindent We have $\theta$ as the mean vector of the 4000 posterior draws of ($\mu_\alpha$, $\mu_\beta$, $\Sigma_{11}$, $\Sigma_{12}$, $\Sigma_{22}$) from fitting the data-rich model to data-rich days. $\Upsilon$ is the 5x5 covariance matrix of $\mu_\alpha$, $\mu_\beta$, $\Sigma_{11}$, $\Sigma_{12}$ and $\Sigma_{22}$, constructed using their 4000 posterior draws from the data-rich model. We fit the data-poor model to all data-poor days again using Stan and the NUTS MCMC algorithm. \\

\subsection{Reparameterisation}\label{sec:reparameterisation}
When fitting the data-rich and data-poor models to actual observations, the MCMC sampler was confronted with an abundance of divergent transitions. To remove these transitions, we reparameterise our models into effectively equivalent mathematical representations that present a simpler posterior geometry \cite{Omiros2007,Stan}. We do so making use of the Cholesky decompositions of our covariance matrices \cite{Betancourt2013}. \\

In order to decrease fitting time of the data-rich model, we replace the per-pixel precisions returned with the TROPOMI data products with daily averages, defined by 

\begin{align}
    \sigma_{\mathrm{C}t} &= \frac{1}{N}\sum_{i=1}^{N} \sigma_{\mathrm{C}i,t} \\
    \sigma_{\mathrm{N}t} &= \frac{1}{N}\sum_{i=1}^{N} \sigma_{\mathrm{N}i,t}.
\end{align}

This yields a new likelihood equation given by 

\begin{equation}
    \mathrm{CH}_{4i}^{\mathrm{obs}} \sim \mathcal{N}\left(\alpha_t + \beta_t\,\mathrm{NO}_{2i}^{\mathrm{obs}},\,\beta_t^2\,\sigma_{\mathrm{N}t}^2 + \gamma_t^2 + \sigma_{\mathrm{C}t}^2\right)\label{eq:reparameterised_likelihood}.
\end{equation}

\noindent In order to determine how this affects the predictive accuracy of the data-rich model, we fit a data-rich model using per-pixel precisions and a data-rich model using daily averages of precision to data-rich days for the month of January 2019, and estimate the difference between the two models' expected log pointwise predictive density (ELPD) using the widely available information criterion (WAIC) \cite{Watanabe2010,Vehtari2017}. 
We estimate that the difference in ELPD of the two models in this case is $6.92 \pm 7.09$. As the difference in ELPD is within a standard error of zero we continue using daily averages of TROPOMI pixel precision in order to decrease the fitting time of our data-rich and data-poor models.\\

\subsection{Predictions}\label{sec:predictions}
We can predict $\mathrm{CH}_4^{\mathrm{pred}}$ and precision $\sigma_{\mathrm{C}}^{\mathrm{pred}}$ from a fitted model using an observation of nitrogen dioxide $\mathrm{NO}_2$ and its associated precision $\sigma_\mathrm{N}$ as input. To obtain predictions, we first sample $K=1000$ potential values of $\mathrm{CH}_4$ from the model values via $\mathrm{CH}_{4,k} \sim \mathcal{N}\left(\alpha_{t,k} + \beta_{t,k}\,\mathrm{NO}_2,\,\beta_{t,k}^2\,\sigma_{\mathrm{N}}^2 + \gamma_{t,k}^2\right)$ where the subscript $k$ denotes a random draw with replacement of a set of parameter values from one of the 4000 sets in the posterior chain. We then have $\mathrm{CH}_4^{\mathrm{pred}} = \frac{1}{K}\sum_{k=1}^K\mathrm{CH}_{4,k}$ and $\sigma_{\mathrm{C}}^{\mathrm{pred}} = \sqrt{\frac{\sum_{k=1}^K\left(\mathrm{CH}_{4,k} -\mathrm{CH}_4^{\mathrm{pred}} \right)^2}{K}}$. If predictions CH$_4^{\mathrm{pred}}$ were to be used in an inversion analysis, $\sigma_\mathrm{C}^{\mathrm{pred}}$ would be the error associated to CH$_4^{\mathrm{pred}}$.

\subsection{Dropout testing}\label{sec:dropout_validation}
We perform dropout testing in order test the predictive ability of our model. For each data-rich day $t$ we ignore a random selection of 20\% of the co-located observations of methane and nitrogen dioxide and fit the model to the remaining 80\% of the data. We use Stan to encode our model and sample the posterior using NUTS, with four independent chains each with 500 burn-in draws and 1000 retained sampled draws for a total of 4000 draws from the posterior. \\

After the data-rich model has been fit to the retained 80\% of the data, we can compare observed values of methane from the held-out subset of data to predictions from the model and summarise how well the model has fit each day using a reduced chi-squared statistic. The reduced chi-squared statistic on day $t$ is calculated via  

\begin{align}
    \chi_{\nu,t}^2 &= \frac{\chi^2_t}{\nu_t} \\
    \chi^2_t &= \sum_{i=1}^{N}\frac{\left(\mathrm{CH}_{4,i}^{\mathrm{obs}} - \mathrm{CH}_{4,i}^{\mathrm{pred}}\right)^2}{\sigma_{\mathrm{C},i}^2 + \sigma_{\mathrm{C},i}^{\mathrm{pred^{2}}}} \\
    \nu_t &= N - 8,
\end{align}

\noindent where again $N$ is the number of co-located observations of methane and nitrogen dioxide in the study region on day $t$.  After examining the resulting distribution of calculated values of $\chi^2_{\nu,t}$, we fit the model to the entire set of observations in the year 2019, without withholding any subset of the data. This fitted model is the model that we use for predicting values of methane for our final results. \\

\subsection{VIIRS}\label{sec:viirs}
We next retrieved VIIRS Nightfire observations of lit methane flare stacks for each day in 2019. Nightfire datasets provide the location coordinates, estimated temperature and source size of identified flares \cite{Elvidge2013}. We reduce the daily datasets down to flare stacks identified within the study region regardless of estimated temperature. Daily tallies of lit flare stacks were collected for eventual comparison against time series of daily fitted model parameters.\\

\subsection{ERA5}\label{sec:era5}
The TROPOMI $\mathrm{CH}_4$ Level 2 Data Product provides the dry-air column average mixing ratio of methane in units of ppbv, defined as the ratio between the column density of methane to the column density of dry air at each pixel location \cite{CH4_ATBD}. To convert the dry-air column average mixing ratio of methane back to a column density, we simply multiply by the value of the dry air column density at the pixel location, which we calculate ourselves from ERA5 data. Although values of dry air column density are supplied at TROPOMI pixel locations where a retrieval was successfully performed, we calculate our own grid from ERA5 data in order to avoid any discrepancy between values of dry air column density at the locations of ``original" observed methane pixel values and the locations of model predictions. \\

We began by retrieving hourly spatial grids of surface pressure $P_{\mathrm{S}}$ and vertical column of water vapour $VC_{\mathrm{H}_2\mathrm{O}}$ that cover the entire extent of the study region for each day in 2019 \cite{ERA5}. $P_{S}$ is given in [Pa] and $VC_{\mathrm{H}_2\mathrm{O}}$ given in [kg m$^{-2}$]. From the value $P_{\mathrm{S}}$ we can calculate the total column density of air $VC_{\mathrm{air}}$ via $VC_{\mathrm{air}} = \frac{P_{\mathrm{S}}}{\mathrm{g}}$ where $\mathrm{g} = 9.8$ m s$^{-2}$ is gravity. We then calculate the total vertical column of dry air $VC_{\mathrm{dry}\;\mathrm{air}}$ via $VC_{\mathrm{dry}\;\mathrm{air}} =  VC_{\mathrm{air}} - VC_{\mathrm{H}_2\mathrm{O}}$. \\

\noindent When calculating $VC_{\mathrm{dry}\;\mathrm{air}}$ at the location of either an observation or model prediction of methane, we interpolate the hourly spatial grids of ERA5 data to the grid of TROPOMI methane observations, and then linearly interpolate in time between the two adjacent hourly grids according to the pixel scanline \cite{CH4_ATBD}.\\

\subsection{Calculation of methane loading}\label{sec:calculation_of_ch4_loading}
We were interested in observing how the application of our predictive algorithm altered the amount of observed methane over the Permian Basin for the year 2019. By virtue of observing more spatial area, the algorithm would result in an increase of total observed mass of methane To counteract this effect, we subtract a nominal background level of methane from each pixel and then describe how the above-background mass of methane in the study region changes after the application of our predictive algorithm. We linearly interpolate the global monthly marine mean surface value of methane as a far-field reference background, provided by the Global Monitoring Laboratory (GML) at the National Oceanic and Atmospheric Association (NOAA) \cite{NOAA}. \\

The number of moles of methane above the GML/NOAA background contained in a pixel is calculated via $\mathrm{mol}\;\mathrm{CH}_4 = \left(\mathrm{CH}_4 - \mathrm{B}\right) \times 10^9 \times VC_{\mathrm{dry}\;\mathrm{air}} \times A$ where $A$ is the pixel area in square metres, B is the mean global surface value of methane in ppbv, and $\mathrm{CH}_4$ is either the TROPOMI-observed or model-predicted pixel value of $\mathrm{CH}_4$ in ppbv. We then convert from moles to tonnes by multiplying by the molar mass of methane and scaling into tonnes. The precision on the moles of above-background methane contained in a pixel is calculated via $\sigma_{\mathrm{mol}\;\mathrm{CH}_4} = \mathrm{mol}\;\mathrm{CH}_4 \sqrt{\left(\frac{\sqrt{\sigma_{\mathrm{C}}^2 + \sigma_{\mathrm{B}}^2}}{\mathrm{CH}_4 - \mathrm{B}}\right)^2 + \left(\frac{\mathrm{RMSE}_{VC}}{VC_{\mathrm{dry\;air}}}\right)^2}$ where $\sigma_{\mathrm{C}}$ is the standard deviation uncertainty on the value of $\mathrm{CH}_4$ and $\sigma_{\mathrm{B}}$ is the standard deviation uncertainty on the value of B. We take $\mathrm{RMSE}_{VC}$ to be the root mean squared error on the residuals between our ERA5-calculated value of $VC_{\mathrm{dry\;air}}$ and the value returned with the TROPOMI Level 2 $\mathrm{CH}_4$ Data Product at all pixel locations possible on all data-rich days in the year 2019. \\

We calculate the total methane mass above background on a given day in the study region via $\mathrm{total\;mol\;CH}_4 = \sum_{i=1}^{N}\mathrm{mol\;CH}_{4i}$ where $N$ is the number of pixels in the study region where we have either $\mathrm{CH}_4^{\mathrm{obs}}$ or $\mathrm{CH}_4^{\mathrm{pred}}$. In order to obtain a conservative estimate, we only include predicted values of methane $\mathrm{CH}_4^{\mathrm{pred}}$ in the summation that are predicted from values of $\mathrm{NO_2}$ with $\mathrm{NO}_2/\sigma_{N}>2$. The precision of total mol CH$_4$ is given by $\sigma_{\mathrm{total\;mol\;CH}_4} = \sqrt{\sum_{i=1}^{N}\left(\sigma_{\mathrm{mol\;CH}_{4},i}\right)^2} \label{eq:total_mol_precision}$.\\

\newpage

\begin{figure}
\begin{center}
\makebox[0pt]{\begin{minipage}{18.288cm}%
\subfloat{%
\includegraphics[width=18.288cm,keepaspectratio]{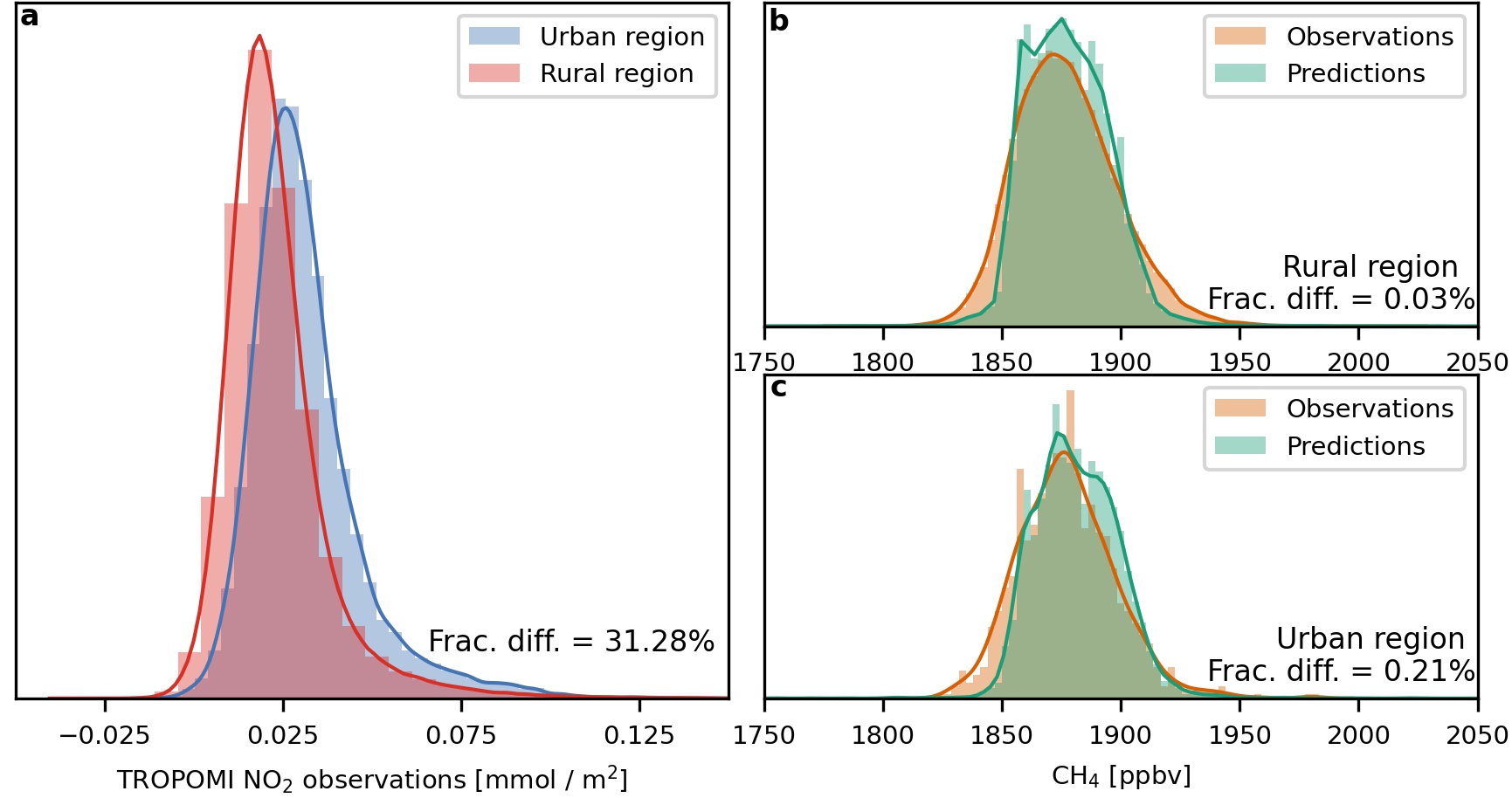}}
\caption{A comparison of distributions of observations over the study region for the year 2019, segregated according to location within the study region. \textbf{a} Distributions of nitrogen dioxide observed over the urban sub-region and the rural surroundings. \textbf{b} Observations and predictions of methane over the rural region made from the fitted model. \textbf{c} Observations and predictions of methane over the urban region made from the fitted model.} \label{fig:city_influence}
\end{minipage}}
\end{center}
\end{figure}

\begin{figure}
\begin{center}
\makebox[0pt]{\begin{minipage}{18.288cm}%
\subfloat{%
\includegraphics[width=18.288cm,keepaspectratio]{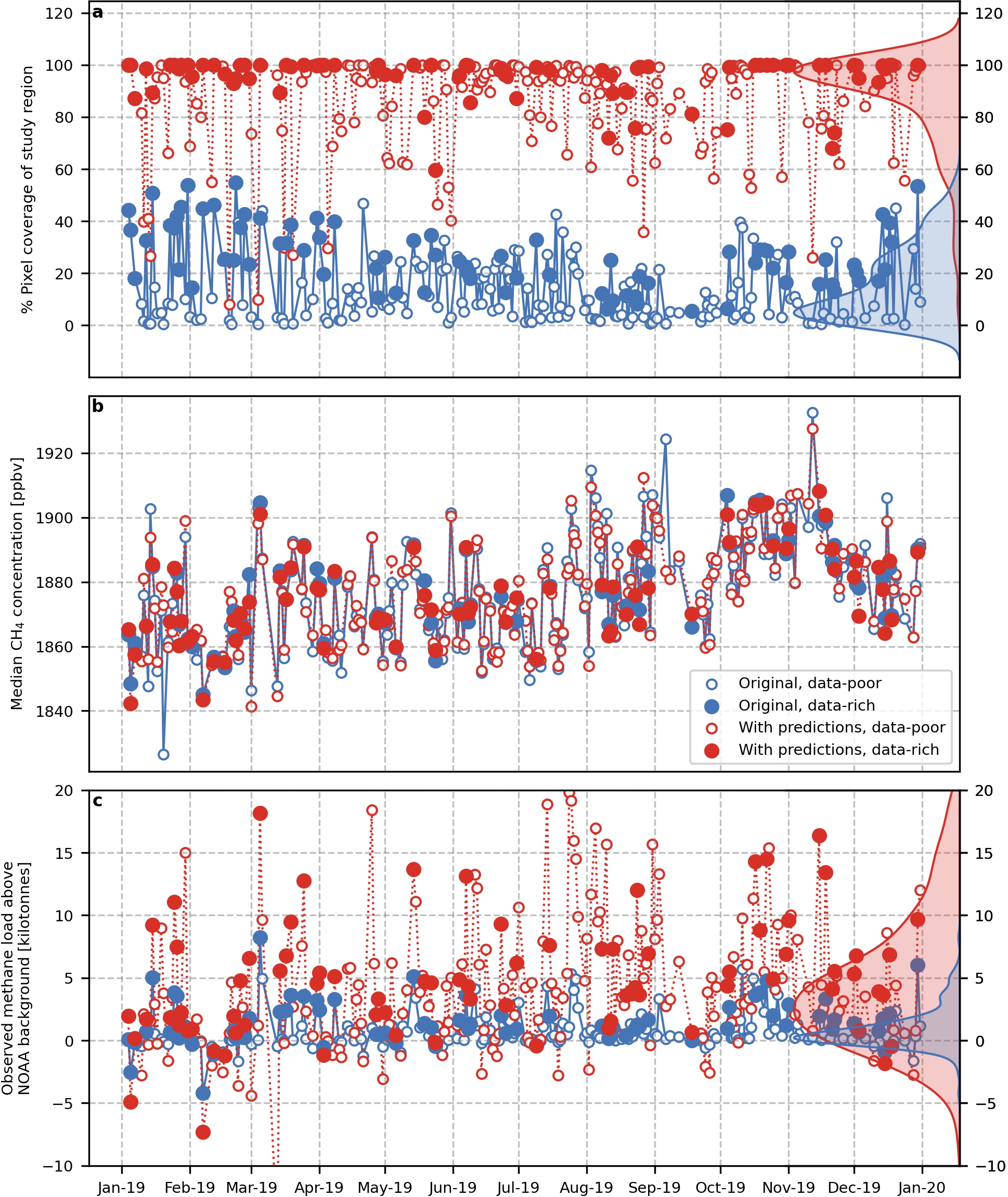}}
\caption{(Continued on the following page.)}\label{fig:5}
\end{minipage}}
\end{center}
\end{figure}

\begin{figure}[t]
  \contcaption{An examination of the effects of including model predictions of methane columns in addition to original TROPOMI observations. \textbf{a-c} In each panel we plot two time series. Plotted in blue are quantities calculated purely from TROPOMI observations that pass the recommended quality assurance threshold. Plotted in red are the same quantities but calculated with the inclusion of predictions of methane concentration from the fitted model at pixel locations in the study region where direct TROPOMI observations of nitrogen dioxide are available but no direct observations of methane are available. For both time series in each panel, full circles indicate data-rich days and open circles indicate data-poor days. \textbf{a} Percentage of usable pixels in the study region, demonstrating that the application of the predictive algorithm augments spatial coverage to nearly 100\% of the study region on most days. \textbf{b} Median observed pixel value in the study region, demonstrating that the inclusion of predictions does not skew the median pixel value in the study region to higher or lower values away from the original median observed pixel value. \textbf{c} Total observed above-background mass of methane over the study region (in reference to the NOAA background). We calculate uncertainty on the quantity plotted in \textbf{c}, but error bars would so narrow as to not be visible when plotted on this scale. Panel \textbf{c} demonstrates that the inclusion of predictions could potentially account for extra kilotonnes of excess methane over the Permian Basin that would nominally be unobserved.}
\end{figure}

\newpage

\end{document}